\documentclass[12pt]{iopart}

\usepackage{graphicx}
\usepackage{iopams}
\usepackage{amssymb}

  \expandafter\let\csname equation*\endcsname\relax
  \expandafter\let\csname endequation*\endcsname\relax
  
\usepackage{amsmath}
\usepackage{bm}
\usepackage{color}
\usepackage[english]{babel}

\newcommand{\ket}[1]{{|{#1}\!\!>}}
\newcommand{\bra}[1]{{<\!\!{#1}|}}
\newcommand{\ip}[2]{{<\!\!{#1}|{#2}\!\!>}}

\begin{document}

\title{Unit Quaternions and the Bloch Sphere}

\author{K B Wharton and D Koch\footnote{Present address: The Graduate Center, CUNY, New York NY 10016.}}

 \address{San Jos\'e State University, Department of Physics and Astronomy, San Jos\'e, CA 95192-0106}
 \ead{kenneth.wharton@sjsu.edu}

\begin{abstract}
The spinor representation of spin-1/2 states can equally well be mapped to a single unit quaternion, yielding a new perspective despite the equivalent mathematics.  This paper first demonstrates a useable map that allows Bloch-sphere rotations to be represented as quaternionic multiplications, simplifying the form of the dynamical equations.  Left-multiplications generally correspond to non-unitary transformations, providing a simpler (essentially classical) analysis of time-reversal.  But the quaternion viewpoint also reveals a surprisingly large broken symmetry, as well as a potential way to restore it, via a natural expansion of the state space that has parallels to second order fermions.  This expansion to ``second order qubits'' would imply either a larger gauge freedom or a natural space of hidden variables.

\end{abstract}

\maketitle

\setlength{\baselineskip}{1.2\baselineskip} 

\section{Introduction}

The unit quaternions form a group that is isomorphic to SU(2), and therefore they have the ideal mathematical structure to represent (pure) spin-1/2 quantum states, or qubits.  But while a unit quaternion $\bm{q}$ is effectively a point on a 3-sphere, a qubit $\psi$ is often represented as a point on a 2-sphere (the Bloch sphere).  Such dimensional reduction results from ignoring the global phase of the spinor $\ket{\chi}$, dropping to a projective Hilbert space where $\ket{\chi}$ and $exp(i\alpha)\ket{\chi}$ correspond to the same qubit $\psi$ (in this case, a Hopf fibration \cite{Penrose,Urbantke}).  This paper examines certain symmetries and natural operations (evident on the full 3-sphere) that have been obscured by this usual reduction; after all, a 3-sphere has a different global geometry than does a circle mapped to every point of a 2-sphere. 

Despite widespread agreement that the SU(2) symmetries of the 3-sphere are more applicable to qubits than are 2-sphere symmetries, the project of analyzing qubits on the full 3-sphere has been generally neglected.  The likely reason is that such an analysis might imply that the global phase has some physical meaning, against conventional wisdom.  To avoid this potential conclusion the global phase is typically removed at the outset.  But the topological mismatch noted above means there is no continuous way to remove this phase from all points on the 3-sphere.  This issue might be seen as a reason to at least temporarily retain the global phase when analyzing spin-1/2 states or equivalent two-level quantum systems.

As further motivation, note that geometric (Berry) phases \cite{Berry} are routinely measured in the laboratory, in seeming contradiction to the orthodox position that global phases are irrelevant.  The typical response here is to deny that single-particle Hilbert spaces are appropriate for measuring relative phases, but nevertheless such phases \textit{can} be computed in a single-spinor framework (see Section 5.2 for further discussion).  This paper takes the position that the predictions of quantum theory would be the same whether global phase is a meaningless gauge or an unknown hidden variable, and the latter possibility is enough to motivate this line of research.

Even if global phases are canonically meaningless, they still can be important to research that strives to extend and/or explicate quantum theory.  Several independent researchers have hit upon using the global phase as a natural hidden variable with a role in probability distributions \cite{Pearle,KGWE,Harrison}, and having a richer single-qubit structure may be useful for ongoing efforts to explain quantum probabilities in terms of natural hidden variables \cite{Argaman,WhartonInfo}.  But for any such work it is important to look at the full 3-sphere, to avoid defining a hidden variable in terms of a globally ill-defined parameter.

For those readers unconcerned with such foundational questions, one can still motivate the 3-sphere viewpoint where it is mathematically advantageous to represent and manipulate spinors in quaternionic form (even if the global phase is eventually discarded).  These applications are developed in the next two sections.  Section 2 defines a useable map between spinors and unit quaternions that conveniently provides a direct quaternion-to-Bloch-sphere mapping.  Right- and left- quaternionic multiplications are shown to correspond to rotations on the Bloch sphere, with particularly surprising results for left-multiplications.  After developing dynamics in Section 3, one immediate application is a dramatic simplification of issues related to time-reversal.  Specifically, one can time-reverse an arbitrary spin-1/2 state via a simple left-multiplication, without either complex-conjugating the state vector or including such a conjugation as part of a time-reversal operator.  This permits a straightforward, classical-style analysis of the time-symmetry of the Schr\"odinger-Pauli equation.

Section 4 looks at further ramifications motivated by this 3-sphere viewpoint.  It is shown that a (seemingly) necessary symmetry-breaking in the Schr\"odinger-Pauli equation (the sign of $i$) looks quite unnatural when framed in terms of quaternions.  This broken symmetry is related to the choice of \textit{which} Hopf fibration one uses to reduce the 3-sphere to the quantum state space.  In section 4.3, restoring this symmetry motivates an alternative second-order dynamical equation, encoding standard dynamics but containing a richer hidden structure.  The extra parameters naturally encode the choice of Hopf fibration, avoiding the broken symmetry while maintaining a clear connection to ordinary quantum states.  Combined with the global phase, these new parameters comprise either a larger gauge symmetry or a natural space of hidden variables.  Section 5 then discusses and expands upon all of these results.

Surveying the literature, the closest analog to the analysis in Section 2 concerns transformations of plane-wave electromagnetic signals, translated into four-dimensional Euclidean space via an extended Jones calculus. \cite{Karlsson}  (That approach used normalized real 4-vectors instead of unit quaternions, but this is not an essential difference.)  While it seems doubtful that the novel digital signaling applications motivated by that research might be applicable in a quantum context, it nevertheless demonstrates that a quaternionic viewpoint can yield new perspectives on a well-understood system.

It has also been noted that the standard mathematics for spinors looks awkward when expressed in quaternionic form, most explicitly in work by Adler \cite{Adler}.  In this prior work, Adler focuses on the complex inner product, and proposes a quaternionic replacement while leaving the dynamics unchanged.  Such a step has the effect of halving the state space, and motivates the field of quaternionic quantum mechanics \cite{Adler2}.  But apart from the initial motivation, it should be noted that the present paper does not follow this path in any way.  Far from extending the traditional machinery of quantum mechanics into the domain of quaternionic inner products, this work simply explores the evident symmetries of the 3-sphere, and tries to preserve such symmetries while maintaining a map to standard spin-1/2 quantum states.  It turns out that this goal can best be accomplished via a dramatically \textit{enlarged} state space; a one-to-many mapping from qubits to quaternions.

\section{Quaternionic Qubits}

\subsection{A Spinor-Quaternion Map}	
	
	A qubit can be represented by any point along the surface of the Bloch sphere, with the north and south poles corresponding to the pure states $\ket{0}$ and $\ket{1}$ respectively, as shown in Figure 1.  (Qubits here are assumed to be pure; a later discussion of mixed states will be framed in terms of distributions over pure states, never as points inside the Bloch sphere.)  For a given point $(\theta,\phi)$ on the sphere (in usual spherical coordinates), the corresponding qubit is defined by 
\begin{equation}
\label{eq:qubit}
\psi =e^{-i\phi/2} cos(\theta/2)\hspace{1mm} |0> + e^{i\phi/2}sin(\theta/2)\hspace{1mm} |1>.
\end{equation}
As noted above, the global phase is not encoded in a qubit, so $\psi$ and $exp(i\alpha)\psi$ correspond to the same physical state.

The distinction between a spinor and a qubit, as used in this paper, is that spinors distinguish between such global phases.  Here a spinor is defined as $\ket{\chi}={a \choose b}$, with $a,b\in\mathbb{C}$ and imposed normalization $\ip{\chi}{\chi}=1$.  Multiplying $\ket{\chi}$ by $exp(i\alpha)$ results in a different spinor, albeit one that corresponds to the same qubit.  It is crucial to note that there is not a globally-unique way to decompose $\ket{\chi}$ into the three angles $(\theta,\phi,\alpha)$, where the first two represent the location of the qubit on the Bloch sphere.  For example, attempting
\begin{equation}
\label{eq:3phase}
\ket{\chi} \hspace{2mm}= e^{i\alpha}\begin{pmatrix}
                cos(\frac{\theta}{2}) e^{-i\frac{\phi}{2}} \\
                sin(\frac{\theta}{2}) e^{i\frac{\phi}{2}} 
        \end{pmatrix} 
\end{equation}
results in a coordinate singularity for qubits on the z-axis, leading to many possible values of $\alpha$.  This reflects the fact that $\ket{\chi}$ naturally represents a point on a 3-sphere, and the global geometry of a 3-sphere is not simply a phased 2-sphere.  And if $\alpha$ cannot be globally defined, it cannot be neatly removed without consequences.

This point is clearer when the spinor is rewritten as a quaternion.  There are many ways to accomplish this, but an obvious choice is the invertible map $M_i\!:\! \ket{\chi} \to \bm{q}$ defined by $\bm{q}=a+b\bm{j}$, where $\bm{q}\in\mathbb{H}$.  (A short primer on quaternions can be found in the Appendix.)  Explicitly, this map reads 
\begin{equation}
\label{eq:qdef}
M_i[\ket{\chi}]=\bm{q}=Re(a)+\bm{i}Im(a)+\bm{j}Re(b)+\bm{k}Im(b).
\end{equation}
\begin{figure}[h]
\includegraphics[width=10cm]{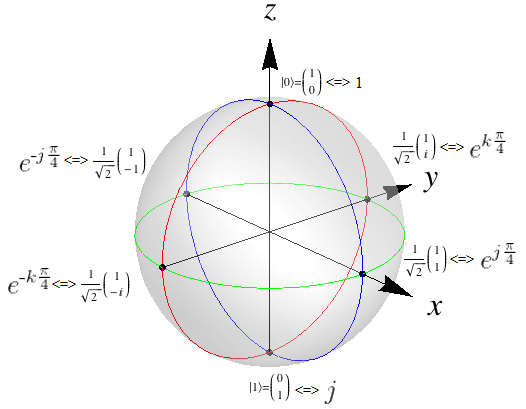}
\caption{Six representative spinors are shown on the Bloch sphere, along with their quaternion equivalent under the map $M_i$.  The states can be (left)-multiplied by a global phase term $exp(\bm{i}\alpha)$, so there are many spinors (and quaternions) at a given point on the sphere.  For example, $\bm{q}\!=\!\bm{i}$ is also at $\ket{0}$, and $\bm{q}\!=\!\bm{k}$ is also at $\ket{1}$.}
\end{figure}

Normalization is enforced by restricting $\bm{q}$ to unit quaternions, $|\bm{q}|^2=1$.  From this it should be evident that the space of all unit quaternions lies on a unit 3-sphere, and so, therefore, does the space of all normalized spinors.

The previous point concerning the ambiguity of $\alpha$ can also be made clearer in a quaternionic context.  Under the map $M_i$, the quaternionic version of (\ref{eq:3phase}) is

\begin{equation}
\label{eq:q3phase}
\bm{q} =  e^{\bm{i}\alpha}e^{\bm{j}\frac{\theta}{2}} e^{-\bm{i}\frac{\phi}{2}},
\end{equation}
from which it is evident that if $\theta=0$, only the combination $(\alpha-\phi/2)$ can be assigned a unique value.  Despite this ambiguity, (\ref{eq:q3phase}) can always be used to find the corresponding Bloch sphere unit vector $\hat{q}$ in spherical coordinates.  But if $\bm{q}$ is not already of the form in (\ref{eq:q3phase}), it would seem to be easier to find $\hat{q}$ by inverting the map $M_i$ (\ref{eq:qdef}) and using standard spinor analysis (which involves discarding the global phase). 

A more elegant method for finding the Bloch sphere unit vector $\hat{q}$ \textit{without} passing through the spinor representation is to generate a unit pure quaternion $\hat{\bm{q}}$ (with no real component) and then map $\hat{\bm{q}}$ directly to $\hat{q}$ in Cartesian coordinates.  Assuming the map $M_i$, this can be done via
\begin{eqnarray}
\label{eq:qhat}
\hat{\bm{q}} =\bar{\bm{q}}\bm{iq}, \\
\label{eq:fdef}
\hat{q}=f[\hat{\bm{q}}]\equiv\hat{\bm{q}}_k\hat{x}-\hat{\bm{q}}_j\hat{y}+\hat{\bm{q}}_i\hat{z}.
\end{eqnarray}

Here $\hat{\bm{q}}_i$ is the $\bm{i}$-component of $\hat{\bm{q}}$, \textit{etc.}, and this last equation is easily invertible, $\hat{\bm{q}}=f^{-1}[{\hat{q}}]$, given the Cartesian components of $\hat{q}$.  The former equation (\ref{eq:qhat}), however, is not invertible; inserting the form of $\bm{q}$ from (\ref{eq:q3phase}) into (\ref{eq:qhat}), one finds that the global phase $\alpha$ always disappears exactly.  

But despite the mathematical elimination of $\alpha$ when mapping to the Bloch sphere, this quaternionic perspective still permits a geometrical interpretation of the global phase.  This is because Eqn. (\ref{eq:qhat}) is known to represent a rotation $\bm{i}\to\hat{\bm{q}}$ on the 2-sphere of unit pure quaternions (as further discussed in the Appendix).  The different global phases, then, apparently correspond to \textit{different} rotations that will take $\bm{i}$ to the same $\hat{\bm{q}}$. 

These rotations can also be mapped on the Bloch sphere itself.  First, write $\bm{q}$ in the most natural form of an arbitrary unit quaternion;
\begin{equation}
\label{qandw}
\bm{q} = e^{\hat{\bm{w}} \beta},
\end{equation}
where $\hat{\bm{w}}$ is another unit \textit{pure} quaternion, and $\beta$ is an angle.  To interpret (\ref{eq:qhat}) as a rotation on the Bloch sphere, simply map all of the pure quaternions to the Bloch sphere using Eqn (\ref{eq:fdef}); $\hat{q}=f[\hat{\bm{q}}]$, $\hat{z}=f[{\bm{i}}]$, $\hat{w}=f[\hat{\bm{w}}]$.  Eqn (\ref{eq:qhat}) then indicates that the Bloch sphere vector $\hat{q}$ can be found by rotating the vector $\hat{z}$ by an angle $-2\beta$ around the axis $\hat{w}$.  Just as there many rotations that will take one vector into another, there are many quaternions that correspond to any given vector $\hat{q}$.  

This implies that the most natural reading of the spin-1/2 state in the form of the quaternion $\bm{q}$ is not a mere vector on a 2-sphere, but rather as a \textit{rotation} on a 2-sphere. This rotation can be \textit{used} to generate a particular vector $\hat{q}$, but it also contains more information not available in $\hat{q}$, such as the angle $\beta$.  This angle is distinct from the global phase $\alpha$ (as the latter cannot be precisely defined in a global manner).

\subsection{Right Multiplication}

For a spinor represented as a unit vector on the Bloch sphere, a rotation of that vector by an angle $\gamma$ around an arbitrary axis $\hat{n}$ can be achieved by an operation of the complex matrix:
\begin{equation}
\label{eq:Rn}
\bm{R}_{\hat{n}}(\gamma) = cos(\frac{\gamma}{2}) I - i \, sin( \frac{\gamma}{2} ) \hat{n} \cdot \vec{\sigma}
\end{equation}
Here, $\vec{\sigma}$ is the usual vector of Pauli matrices, defined in the Appendix.  

There is a simple correspondence between $\bm{R}_{\hat{n}}(\gamma)$ and an exponential quaternion, due to the strict parallel between $i\vec{\sigma}$ and the three imaginary quaternions $\bm{i,j,k}$.  (See the Appendix and Table 1 for further details on this point.)  Assuming the map $M_i[\chi]\!\!=\!\!\bm{q}$ defined above, a right multiplication by $exp({-\bm{k}\gamma/2})$, $exp({\bm{j}\gamma/2})$, or $exp({-\bm{i}\gamma/2})$ on a unit quaternion $\bm{q}$ rotates the corresponding Bloch sphere vector $\hat{q}$ by an angle $\gamma$ around the positive $\hat{x}$, $\hat{y}$, or $\hat{z}$ axes (respectively).   

More generally, a right multiplication by $exp({ -\hat{\bm{n}} \gamma/2 })$ effects a rotation of an angle $\gamma$ around the arbitrary axis $\hat{n}=f[\hat{\bm{n}}]$ such that:
\begin{equation}
 \label{eq:RM}
M_i[\bm{R}_{\hat{n}}(\gamma) \ket{\chi}] = \bm{q} e^{ -\hat{\bm{n}} \frac{\gamma}{2}} .
\end{equation}
This simple relationship can also be seen from Eqn (\ref{eq:qhat}); as $\bm{q}\to\bm{q}\,\,exp(-\hat{\bm{n}} \gamma/2)$, one finds $\hat{\bm{q}}\to exp(\hat{\bm{n}} \gamma/2)\,\hat{\bm{q}}\,\,exp(-\hat{\bm{n}} \gamma/2)$.  Again, this is a rotation of $\hat{\bm{q}}$ on the 2-sphere of unit pure quaternions, which can be mapped to the Bloch sphere via (\ref{eq:fdef}).

As every unit quaternion can be written in the form $exp({ -\hat{\bm{n}} \gamma/2 })$, and as we are only interested in transformations that keep $\bm{q}$ normalized, there are no other right-multiplications to consider.  Table 1 lists some useful special rotations, corresponding to quantum gates (assuming the use of the $M_i$ map from spinors to quaternions).

\begin{table}
\caption{\label{tab1} Some common single-qubit gates are presented in terms of a right-multiplied quaternion (assuming the use of the map $M_i$).  For $\pm\pi$-rotations, the two possible directions yield a different sign outcome.}
\begin{indented}
    \item[] \begin{tabular}{| c |  c | c | p{6cm} | }
    \hline
    Gate  & Matrix Operator & Equivalent Quaternion   \\
    & & (Right Multiplication)\\ \hline
    Pauli X-Gate & $ \pm i \begin{bmatrix}
0 \hspace{2mm} 1 \\
1 \hspace{2mm} 0
\end{bmatrix}
 $ & { \large $e^{\pm\bm{k} \frac{\pi}{2}}=\pm \bm{k}$}  \\ \hline
    Pauli Y-Gate & $ \mp i \begin{bmatrix}
0 \hspace{2mm} $-$i \\
i \hspace{2mm} 0
\end{bmatrix}
 $ & { \large $e^{\pm\bm{j} \frac{\pi}{2}}=\pm \bm{j}$ } \\  \hline
     Pauli Z-Gate & $ \pm i \begin{bmatrix}
1 \hspace{2mm} 0 \\
0 \hspace{2mm} $-$1
\end{bmatrix}
 $ & { \large $e^{\pm\bm{i} \frac{\pi}{2}}=\pm \bm{i}$ }  \\ \hline
 Phase Shift Gate & $ \begin{bmatrix}
e^{-i \theta/2} \hspace{2mm} 0 \\
\hspace{2mm} 0 \hspace{6mm} e^{i \theta/2}
\end{bmatrix}
 $ & { \Large e$^{-\bm{i} \frac{\theta}{2}}$ } \\ \hline
Hadamard Gate & $  \dfrac{\pm i}{\sqrt{2}} \begin{bmatrix}
1 \hspace{3mm} 1 \\
1 \hspace{2mm} $-$1
\end{bmatrix}
 $ & { \large e$^{\pm\frac{\bm{i}+\bm{k}}{\sqrt{2}} \frac{\pi}{2}}=\pm\frac{\bm{i}+\bm{k}}{\sqrt{2}}$ }  \\ \hline
    \end{tabular}
    
\end{indented}
\end{table}

\subsection{Left Multiplications}

	From a quaternion-based viewpoint, one would expect no essential difference for left-multiplication as compared to right-multiplication; given the above results, left-multiplication should merely encode another class of rotations.  But right-multiplications have seemingly spanned the range of possible unitary operators, so a spinor-based perspective might find it surprising that there is another type of transformation at all.  Indeed, in some quaternion-based approaches to quantum spin \cite{Avron}, left-multiplications have simply been left undefined, despite the clear meaning of such operations in the quaternion algebra.
	
	The resolution of this apparent disagreement lies in the fact that left-multiplications map to non-unitary operators and/or global phase shifts.  (This was recently detailed in an analysis of electromagnetic plane-waves, concerning these same left-isoclinic rotations in 4D Euclidean signal space \cite{Karlsson}.)  Some left-multiplications correspond to anti-unitary operators, making this mathematics particularly useful for analysis of time-reversed spin-1/2 systems.  The connection between quaternions and time-reversal of spin-1/2 states has been known for some time \cite{Avron,Dyson}, but utilizing generic left-multiplications (as opposed to simply a special time-reversal operator) allows for a deeper analysis of the relevant symmetries. 
	
	Except for a measure-zero set of (unitary) phase-change operators and (anti-unitary) time-reversal operators, most of the quaternionic left-multiplications correspond to operators that are merely non-unitary.  A famous theorem by Wigner \cite{Wigner} indicates that such operators must take pure states into mixed states, and such operators have already found significant application in quantum information theory. 
	
	Here a new motivation presents itself, via the quaternion mathematics.  Given that there is no essential difference between quaternionic left- and right-multiplication, every symmetry evident in the right-multiplication sector should have an equally important symmetry in the left-multiplication sector.  (Recall the original motivation of this paper was to see if perhaps the Bloch-sphere viewpoint had obscured particular features evident on the full 3-sphere.)
	
	 As before, we shall assume the map $M_i[\ket{\chi}]\!\!=\!\!\bm{q}$, and enforce the continued normalization of $\bm{q}$ by only multiplying exponential quaternions.  Using the notation $\bm{q}'=\bm{q}_L \bm{q}$, the simplest case is the left multiplication $\bm{q}_L\!=\!exp({\bm{i}\gamma/2})$, which simply changes the global phase of $\bm{q}$.  Eqn  (\ref{eq:qhat}) indicates that this does not lead to any Bloch sphere rotation, because $\hat{\bm{q}}'=\hat{\bm{q}}$.
	 
	 This same equation offers a geometrical interpretation of a general left-multiplication $\bm{q}_L$.  Recall that (\ref{eq:qhat}) implies $\bm{q}$ does not merely encode a Bloch vector $\hat{q}$, but instead a \textit{rotation} of the vector $\hat{z}$, with many different rotations ($\bm{q}$'s) that can take $\hat{z}$ to the same $\hat{q}$ (corresponding to different global phases).  Now, a left multiplication on $\bm{q}$ can be viewed as two consecutive rotations;
	 \begin{equation}
	 \label{eq:qprime}
	 \hat{\bm{q}}'=\bar{\bm{q}} \left( \bar{\bm{q}}_L \bm{i} \bm{q}_L \right) \bm{q}.
	 \end{equation}
In other words, $\bm{q}_L$ serves to \textit{first} rotate $\hat{z}=f[\bm{i}]$, before the rotation encoded by $\bm{q}$ can be executed.  And as rotations do not commute, one cannot calculate $\hat{q}'=f[\hat{\bm{q}}']$ without knowing \textit{which particular rotation} $\bm{q}$ encodes.  Different global phases of the original $\bm{q}$ will therefore lead to a different final Bloch vector $\hat{q}'$, even if $\hat{q}$ and $\bm{q}_L$ are exactly known.  Such a transformation is neither unitary nor anti-unitary, but may be mathematically useful even if it is not physically possible. 

The two exceptions to this ambiguity are when the central term in (\ref{eq:qprime}) is either $\bm{i}$ or $-\bm{i}$.  The former case has been discussed above; this is just a global phase shift.  The latter case can be realized by a left multiplication of (say) $\bm{q}_L=\!\bm{j}$ or $\bm{q}_L=\!\bm{k}$.  The net effect of such a left-multiplication would be a simple minus sign; $\hat{\bm{q}}'=-\hat{\bm{q}}$, inverting the original Bloch sphere vector.  It follows that both $\bm{j}$ and $\bm{k}$ act like the anti-unitary time-reversal operator $\bm{T}$ (when used to left-multiply a quaternion).

If one insists on thinking of $\bm{q}$ as simply encoding a vector $\hat{q}$ on the Bloch sphere (rather than as a rotation), it is still possible to interpret left-multiplications as a rotation of $\hat{q}$ around a sometimes-unknown axis.  A general left multiplication of $\bm{q}$ by $\bm{q}_L=exp({ -\hat{\bm{n}} \gamma/2 })$ is equivalent to rotating the Bloch sphere vector $\hat{q}$ by an angle $\gamma$ around some axis.  But this rotation axis $\hat{r}$ is no longer given by $\hat{n}=f[\hat{\bm{n}}]$.  Indeed, $\hat{r}$ is not even computable from knowledge of the vector $\hat{q}$; it depends on the entirety of $\bm{q}$.  Still, if the global phase $\alpha$ is completely unknown, instead of one particular rotation axis $\hat{r}$, there are instead many possible rotation axes, $\hat{r}(\alpha)$.   These possible axes form a circle on the Bloch sphere, and $\hat{q}$ passes through the center of this circle.  
	
	The result is that $\bm{q}_L$ corresponds to a cone of possible rotation axes (assuming the global phase is unknown).  The angle of this cone can be determined from the relationship
\begin{equation}
\label{eq:cone}
\hat{n} \cdot \hat{z} = \hat{q} \cdot \hat{r}.
\end{equation}
In other words, the angle between the $z$-axis and $\hat{n}$ is the half-angle of the cone produced by the possible values of $\hat{r}$.  In the special case that $\hat{\bm{n}}=\bm{i}$, this cone angle is zero.  In this case, the only possible rotation axis for $\hat{q}$ is $\hat{q}$ itself, or no rotation at all; this corresponds to a global phase change, with no state change.

The other special case is when $\hat{\bm{n}}$ lies in the quaternionic $\bm{j}-\bm{k}$ plane, which means that $\hat{n}$ lies in the $x-y$ plane of the Bloch sphere.  The angle between $\hat{z}$ and $\hat{n}$ is then always $\pi/2$.  In this case, the possible rotation axes $\hat{r}(\alpha)$ form a great circle: the equator corresponding to a pole defined by $\hat{q}$.  A $\gamma=\pi$ rotation around \textit{any} of these axes will send $\hat{q}\to -\hat{q}$, exactly reversing the direction of the Bloch sphere vector, regardless of $\alpha$.  (Again, this corresponds to an anti-unitary operation, $\bm{T}$.)  In general, for the map $M_i[\ket{\chi}]=\bm{q}$, this reversal is equivalent to any left-multiplication of the form
\begin{equation}
\label{eq:Tdef}
M_i[\bm{T}\ket{\chi}]=e^{[\bm{j} cos(\delta) + \bm{k} sin(\delta)]\pi/2} \bm{q}
\end{equation}
for any angle $\delta$.  One convenient left-multiplication of this form is at $\delta=0$, or $\bm{jq}$, which will be used as the time-reversed representation of $\bm{q}$ in the next section.

\section{Dynamics}

\subsection{Spin-1/2 in a magnetic field}

When it comes to the equations that describe the dynamics of a charged spin-1/2 state in a magnetic field, quaternions also provide a useful and simplifying framework.  In a magnetic field $\vec{{B}}(t)$, the standard Schr\"odinger-Pauli equation for $\ket{\chi(t)}$ reads (in spinor form)
\begin{equation}
\label{eq:sspe}
i\hbar \frac{d}{dt} \ket{\chi}= -\gamma \frac{\hbar}{2} \vec{{B}}\cdot\vec{\sigma} \, \ket{\chi}.
\end{equation}
Here $\gamma$ is the gyromagnetic ratio.

Because of the correspondence between the three components of $i\vec{\sigma}$ and the imaginary quaternions $(\bm{i},\bm{j},\bm{k})$, this matrix algebra can be trivially encoded in the quaternionic version of the Schr\"odinger-Pauli equation, which simply reads
\begin{equation}
\label{eq:qspe}
\dot{\bm{q}}=-\bm{q}\bm{b}.
\end{equation}
Here $\bm{b}$ is a pure quaternion defined in terms of the three components of $\vec{{B}}$;
\begin{equation}
\label{eq:bdef}
\bm{b} \equiv  \frac{\gamma}{2} \left( \bm{i} B_z - \bm{j} B_y + \bm{k} B_x \right).
\end{equation}

But these equations are unsatisfactory in that they only describe the geometric phase, and this is not measureable on its own; only the combined dynamic plus geometric phase can be detected.  An inclusion of even the simplest and most fundamental dynamical phase (say, a constant-energy term $exp(-i\omega_0t)$, where the energy $\hbar\omega_0$ might include a rest mass) dramatically changes these equations.   Indeed, in the limit $\vec{\bm{B}}\to0$, this would be the only surviving phase.  

Inclusion of this simplest dynamic phase would appear as an extra term $\hbar\omega_0\chi$ on the right side of (\ref{eq:sspe}).  The corresponding quaternionic equation (\ref{eq:qspe}) is
\begin{equation}
\label{eq:spe}
\dot{\bm{q}}=-\bm{q}\bm{b}-\bm{i}\omega_0\bm{q}.
\end{equation}
Crucially, while $\bm{b}$ enters as a right-multiplication, a quaternionic $\bm{i}$ enters as a left-multiplication.  Somehow, one particular pure quaternion $(\bm{i})$ has been singled out by the dynamics, over $\bm{j}, \bm{k},$ etc.

The source of this asymmetry can be traced back to the original map $M_i$, defined in Section 2; other choices would have resulted in a different final term of (\ref{eq:spe}).  To see this, define a different map $M_v(\chi)\equiv \bm{u}M_i(\chi)$, where $\bm{u}$ is any unit quaternion.  Under this alternate map, one finds a new representation of the spin state $M_v(\ket{\chi})=\bm{q'}$, related to the old representation by $\bm{q'}=\bm{uq}$ (or, equivalently, $\bm{q}=\bar{\bm{u}}\bm{q'}$).  Using this in (\ref{eq:spe}), and left-multiplying by $\bm{u}$, results in
\begin{eqnarray}
\label{eq:vspe}
\dot{\bm{q'}}&=&-\bm{q'}\bm{b}-\hat{\bm{v}}\omega_0\bm{q'},\\
\label{eq:vdef}
\hat{\bm{v}}&\equiv& \bm{ui\bar{u}}.
\end{eqnarray}
Here the unit quaternion $\hat{\bm{v}}$ is guaranteed to be pure. 

One can also consider alternate maps of the form $M_i \bm{r}$, but as detailed in section 2.2, this is merely a rotation of the entire coordinate system, and only changes the map $f$ between the pure quaternions and the Bloch sphere (\ref{eq:fdef}).  (Also, it changes the inverse of this map, which shows up in $\bm{b}$ via (\ref{eq:bdef}).)  The most general map, $\bm{u} M_i \bm{r}$, then, defines both the coordinate system in which the Bloch sphere is embedded and the pure quaternion $\hat{\bm{v}}$ in the dynamical equation (\ref{eq:vspe}). 

\subsection{Application: Time Reversal}

One example of the value of quaternionic equations can be found by examining the issue of time-reversal.  Applied to the standard (\ref{eq:sspe}), one finds a series of decisions requiring substantial care and expertise to get correct: Does the non-unitary time-reversal operator $\bm{T}$ apply to such differential equations, or merely to instantaneous states?  Does the sign of $i$ change along with the sign of $d/dt$, or is conjugation itself a way of effecting $t\to -t$?  Are the Pauli matrices all time-odd, or just the imaginary $\sigma_y$?  Is energy time-odd or time-even?  This section argues that such questions become far more straightforward when applied to (\ref{eq:spe}), and might even provide fresh insights to curious fermionic features such as $\bm{T}^2=-1$.

For the quaternionic equation, it turns out that one can use the same logic as time-reversal in classical physics; no complex conjugations are required.  Namely, one first changes the sign of all the time-odd physical quantities, then changes the sign of $t$ in all of the differential equations, and finally looks to see if the transformed equation has the same form as the original equation.  (If so, one has time-symmetric physical laws.) In the case of  (\ref{eq:spe}), the only time-odd quantities are the magnetic field $\bm{b}$ and the angular momentum encoded by $\bm{q}$.  Importantly, classical energy is time-even (recall, $\omega_0$ might represent a rest mass term, $mc^2/\hbar$), and should not change sign under time reversal.

Section 2.3 showed that one can reverse the direction of any arbitrary spin direction encoded by $\bm{q}$ via $\bm{q}\to\bm{jq}$.  If one also takes $\bm{b}\to -\bm{b}$ and changes the sign of the time-derivative on the left side of  (\ref{eq:spe}), this equation becomes
\begin{equation}
\label{eq:Trev}
-\bm{j}\dot{\bm{q}}=+\bm{j}\bm{q}\bm{b}-\bm{i}\bm{j}\omega_0\bm{q}.
\end{equation}
Another left-multiplication by $\bm{j}$ therefore restores the exact form of (\ref{eq:spe}), because $\bm{jij}=\bm{i}$.  Therefore the Schr\"odinger-Pauli equation is time-symmetric, in precisely the same sense as classical physics.

If one performs this time-reversal twice, and demands the same left-multiplication on $\bm{q}$ for each time-reversal, one has (say) $\bm{q} \to \bm{j}^2\bm{q}=-\bm{q}$, matching the $\bm{T}^2=-1$ operation from ordinary quantum physics.  But from a classical perspective, such an equation appears baffling; surely if one performs two time-reversals on any given history $\bm{q}(t)$, it should be a logical imperative that one recovers the original $\bm{q}(t)$, not a phase-reversed version.  A solution to this quantum-classical disconnect could plausibly lie in the ambiguity of exactly \textit{which} quaternion should be used to implement time-reversal  -- or better yet, a removal of this ambiguity entirely.  

Looking at the more general (\ref{eq:vspe}), one obvious strategy would be to associate time-reversal directly with a change of the sign of $\hat{\bm{v}}$, rather than any particular left-multiplication on $\bm{q}$.  If such a step were meaningful, it would immediately solve the above problems; two time-reversals would then end up back at the original solution, with no extra phase shift.  

Unfortunately, at this point, such a step does not seem to be possible, as $\hat{\bm{v}}$ is only defined in terms of the choice of $\bm{u}$ used in the map $M_v$, and the ambiguity of how to choose $\bm{u}$ remains.  (One could equally well choose $\bm{u}=\pm\bm{j}$ or $\bm{u}=\pm\bm{k}$ to change the sign of $\hat{\bm{v}}$.)  But if one follows through the proposals in the next section, it is possible to extend the dynamics such that $\hat{\bm{v}}$ does indeed have independent physical meaning as a time-odd parameter, raising the possibility that $\bm{T}^2=-1$ could be reconciled with the classical meaning of time-reversal.

\subsection{The Generalized Map}

In Section 3.1, we noted that use of the map $M_i$ singled out $\bm{i}$ as a special quaternion, and a careful reader might have noticed that this was also the case throughout Section 2.  The quaternion $\bm{i}$ corresponds to the unit vector $\hat{z}=f[\bm{i}]$, which played several special roles.  In Section 2.1, $\hat{z}$ was the starting vector for the natural interpretation of $\bm{q}$ as a rotation.  In Section 2.3, the unit vector $\hat{z}$ appears in Eq. (\ref{eq:cone}).  Note that it was a left multiplication of $exp(-\bm{i}\gamma/2)$ that changed the global phase, but any pure quaternion other than $\bm{i}$ in the exponent led to a non-unitary transformation of $\bm{q}$.

It is straightforward to generalize Section 2 to work for any map $M_v(\ket{\chi})\equiv \bm{u}M_i(\ket{\chi})$ between spinors and quaternions.  Inserting $\bm{q}=\bar{\bm{u}}\bm{q'}$ into (\ref{eq:qhat}), and dropping the primes, one finds simply
\begin{equation}
\label{eq:vqhat}
\hat{\bm{q}} =\bar{\bm{q}}\bm{\hat{v}}\bm{q},
\end{equation}
where $\hat{\bm{v}}$ is defined in (\ref{eq:vdef}).  One can continue to use the same map $\hat{q}=f[\hat{\bm{q}}]$ from these pure quaternions to the Bloch sphere defined by (\ref{eq:fdef}) unless one further generalizes the map $M_v$ with a right-multiplication; this would rotate the coordinate system as described at the end of Section 3.1.

Given (\ref{eq:vqhat}), the generalization of quaternion multiplication to the map $M_v(\chi)$ is straightforward.  In this general case $\hat{\bm{v}}$ is the special pure quaternion instead of $\bm{i}$.  Instead of a special unit vector $\hat{z}$, one instead has a special unit vector $\hat{v}=f[\hat{\bm{v}}]$.  So a generic state $\bm{q}$ describes a rotation from $\hat{v}$ to $\hat{q}$, and it is the vector $\hat{v}$ that is first acted upon by a left rotation $\bm{q}_L$.  With this change of $\hat{z}\to\hat{v}$, the results of Section 2 go through for the general map $M_v$.

\section{Expanded Dynamics}

\subsection{A Broken Symmetry}

Moving beyond (\ref{eq:sspe}), the more general Schr\"odinger equation, for any quantum system $\ket{\psi(t)}$, reads
\begin{equation}
\label{eq:sse}
i\hbar \frac{\partial}{\partial t} \ket{\psi}= \bm{H} \ket{\psi},
\end{equation}
where $\bm{H}$ is the Hamiltonian operator, possibly time-varying.  If one treated the wavefunction as purely time-even (or purely time-odd), the analysis normally applied to classical physics equations (as described in the previous section) would reveal a formal time-aysmmetry. This is because $\bm{H}$ represents energy, a time-even quantity, and time-reversal would lead to a different equation, with different solutions $\ket{\phi}$:
\begin{equation}
\label{eq:trse}
-i\hbar \frac{\partial}{\partial t}  \ket{\phi}= \bm{H} \ket{\phi}.
\end{equation} 

This issue has a well-known resolution; if the wavefunction is also complex-conjugated along with sending $t\to -t$ (or more-generally, $\ket{\psi'(t)}=\bm{T}\ket{\psi(-t)}$), then this conjugated state $\ket{\psi'}$ will solve the original (\ref{eq:sse}).

But even with this resolution, a broken symmetry remains; what chooses the sign of $i$ in (\ref{eq:sse}), and why should such a choice be necessary in the first place?  Almost everyone would agree that this choice is a mere convention, and that equivalent physical predictions would have results if Schr\"odinger had picked the opposite sign for $i$ in his original equation.  But the fact that such a choice was necessary in the first place is an indication of a broken symmetry.

Avoiding this choice is possible, but only by going to the second-order (Klein-Gordon) equation;
\begin{equation}
\label{eq:kge}
-\hbar^2\frac{\partial^2}{\partial t^2}  \ket{\kappa}= \bm{H}^2 \ket{\kappa},
\end{equation}
with the general solution $\ket{\kappa}=A\ket{\phi}+B\ket{\psi}$.  (Here $\ket{\phi}$ and $\ket{\psi}$ are not related to each other, giving $\ket{\kappa}$ twice as many free parameters as either $\ket{\phi}$ or $\ket{\psi}$ alone.)  But despite the symmetries and relativity-friendly nature of this equation, it yields solutions with no obvious single-particle interpretation -- in particular, solutions for which $A$ and $B$ are both non-zero.  (For another perspective, see \cite{KGWE}.)  If such solutions are rejected as unphysical, that forces one to reduce the relevant equation down to either (\ref{eq:sse}) or (\ref{eq:trse}), and that choice would seem to be a \textit{necessary} broken symmetry.

\subsection{The Quaternion Viewpoint}

The arguments in the previous subsection do not properly go through for a spin-1/2 system, and this is most clearly seen from the perspective of quaternionic qubits.  For simplicity, first consider the zero-field limit ($\bm{b}\to0$).  Eliminating this magnetic field term from (\ref{eq:spe}) one might seem to still have a (quaternionic) $\bm{i}$ present, but as discussed above, this stems from the choice $M_i$ of how one maps the spinor to the quaternion $\bm{q}=M_i(\chi)$.  A more-general map $M_v$ yields (\ref{eq:vspe}), or in the zero-field case,
\begin{equation}
\label{eq:vspe2}
\dot{\bm{q}}=-\bm{\hat{v}}\omega_0\bm{q}.
\end{equation}
As before, $\omega_0$ could represent a rest mass ($\omega_0=mc^2/\hbar$), and $\hat{\bm{v}}$ is an arbitrary pure unit quaternion.

The usual link between the standard-form Schr\"odinger equation (\ref{eq:sse}) and the time-reversed Schr\"odinger equation (\ref{eq:trse}) carries over to the quaternions.  Specifically, replacing $\hat{\bm{v}}=\bm{i}$ with $\hat{\bm{v}}=-\bm{i}$ is accomplished via the equivalent of time-reversing the spin-state, $\bm{q}\to\bm{jq}$.  But in this form it is clear that $\hat{\bm{v}}=\pm\bm{i}$ are not the only two options; depending on the choice of map $M_v$, $\hat{\bm{v}}$ could be \textit{any} pure unit quaternion, spanning a smoothly-connected 2-sphere of possible values.  Far from being a clearly broken symmetry (as is the disconnected $\pm$ sign on the complex $i$), one might now ask whether this enlarged and connected symmetry must be broken at all.  

Given the above analysis, the symmetry \textit{must} be broken, because one must choose \textit{some} map $M_v$ to interpret $\bm{q}$ and to define $\hat{\bm{v}}$ in (\ref{eq:vspe}) and (\ref{eq:vspe2}).  The freedom of such a definition lies on a 2-sphere, and is larger than the usual U(1) phase freedom; this is less surprising if one notices that there is also the same freedom when choosing a particular Hopf fibration \cite{Thurston}.  Also note that the generator of a transformation between different choices of $\hat{\bm{v}}$ is a left-multiplication, and is therefore nonunitary, as per Section 2.3.  If a gauge is fixed (for example, setting $\hat{\bm{v}}=\bm{i}$ by fiat), then one can ignore this symmetry in the space of left-multiplications and proceed as usual.  However, the question remains whether this symmetry must be broken at all, especially as it is not merely choosing a sign convention for a complex $i$.

\subsection{Restoring the Symmetry}
	
One can avoid breaking this symmetry without ever using a non-unitary transformation, so long as the particular value of $\hat{\bm{v}}$ has a physical meaning and does not appear in (or change) the form of the dynamical equations.  As in the case of the Klein-Gordon equation, this goal can naturally be accomplished by extending (\ref{eq:vspe2}) to the second-order dynamical equations familiar from classical field theory;
\begin{equation}
\label{eq:vkge}
\ddot{\bm{q}}=-\omega_0^2\bm{q}.
\end{equation}
In the case of the Klein-Gordon equation, this is thought to be unacceptable because the larger solution space contains solutions that do not reduce to those of the first-order equation.  But for the special case of qubits, at least, this concern disappears.  So long as one constrains $\bm{q}$ to be a \textit{unit} quaternion, every solution to this equation will also solve (\ref{eq:vspe2}) for some pure unit quaternion $\bm{\hat{v}}$.  \cite{CarlosThesis}\footnote{This statement is not technically correct for the zero-magnetic field case, as there is nothing to break the symmetry between right- and left- multiplications; some solutions to (\ref{eq:vkge}) will instead solve $\dot{\bm{q}}=-\omega_0\bm{q}\bm{\hat{v}}$.  But this caveat goes away for microscopically-varying magnetic fields; all unit-quaternion solutions to (\ref{eq:bkge}) will solve (\ref{eq:vspe}).}  The larger solution space does indeed have new free parameters, but those parameters are the unit quaternion $\bm{u}$ that defines the map $M_v$ and also defines $\bm{\hat{v}}$ via (\ref{eq:vdef}).  If all maps to the standard Schr\"odinger-Pauli dynamics are indistinguishable (as implied by the above discussion), then $\bm{u}$ is either a choice of gauge or a hidden parameter.  

This result is not specific to the zero-field case.  Adding back the magnetic field, the second-order version of (\ref{eq:vspe}) can be found by taking a derivative and eliminating $\hat{\bm{v}}$;
\begin{equation}
\label{eq:bkge}
\ddot{\bm{q}}+2\dot{\bm{q}}\bm{b}+\bm{q}(\bm{b}^2+\dot{\bm{b}}+\omega_0^2)=0.
\end{equation}
This looks a bit more cumbersome than the first-order (\ref{eq:spe}), but notably it also results from a simple real Lagrangian density (see (\ref{eq:2ndL}) below), and yields solutions that exactly map onto the standard Schr\"odinger-Pauli equation. \cite{WLS}  (The unit quaternion condition can also be imposed via $L=0$, and the unit quaternion $\bm{u}$ encodes the new constants of the motion. \cite{CarlosThesis}).  

This second-order equation on a 4-component quaternion raises the question of whether this is an alternate path to Dirac's extension of 2-spinors to 4-spinors.  After all, a first-order differential equation on a 4-component complex spinor has solutions with 8 free real parameters, just like the solutions to (\ref{eq:bkge}).  But at this level the only constraint on a Dirac spinor is an overall normalization, leaving 7 available parameters (6 if one ignores the global phase).  Here, apart from normalization at some reference time $\bm{q}(t\!\!=\!\!0)=\bm{q_0}$, the solution space is constrained by the additional normalization conditions $|\dot{\bm{q}}(t=0)|=0$ and $|\ddot{\bm{q}}(t=0)|=0$, reducing the solution space to 5 free real parameters, counting the phase.  And every one of these solutions can be made to map back to the original spin-1/2 system; there is no room for antimatter solutions, even in this expanded space.   

Of these five parameters, three can be made to correspond to $\bm{q_0}$.  The remaining 2 parameters are encoded in $\dot{\bm{q}}(t=0)$, and are of course time-odd; these can be made to correspond to $\hat{\bm{v}}$, and they determine which map $M_v$ should be used to interpret ${\bm{q}}$.  (There is also a time-odd (dynamic) phase term in $\bm{u}$, but this naturally combines with the time-even (geometric) phase term in $\bm{q}_0$ to determine a single parameter that corresponds to the net global phase.)

The immediate result of this expanded dynamics is that  $\hat{\bm{v}}$ now encodes the time-odd parameters, and its sign should be changed upon time-reversal.  This not only further simplifies the time-reversal analysis of (\ref{eq:vspe}) above, but provides a non-operator technique for time-reversal, such that two time-reversals always exactly cancel.  Further implications of this expanded dynamical equation (\ref{eq:bkge}) will be discussed in Section 5.3.

\section{Discussion}

\subsection{Summary of Basic Results}

Most of the results from the first three sections do not require or imply any new physics; they simply follow from a reversible map $M_i[\ket{\chi}]=\bm{q}$ between spinors and quaternions.  These results will be of the most interest to the widest audience (even those not interested in foundations) and they will be summarized here first.  A discussion of the more speculative implications will follow.

In the traditional spinor representation there is a clear mathematical distinction between a state $\ket{\chi}={a \choose b}$ and a unitary operator that acts on this state as a 2x2 matrix $\bm{U}$.  In addition, there are non-unitary operators that cannot even be written in matrix form, such as the time-reversal operator $\bm{T}$.     

Mapping spinors onto the unit quaternion $\bm{q}=Re(a)+\bm{i}Im(a)+\bm{j}Re(b)+\bm{k}Im(b)$ reframes all of these distinctions, as both states and operators can now be written as unit quaternions.  There are still more operators than states, because the most general linear transformation of $\bm{q}$ requires both left- and right- multiplications, $\bm{q}'=\bm{q}_L\, \bm{q} \,\bm{q}_R$.  Still, it is notable that all unitary operators can be represented as a quaternion right-multiplication only ($\bm{q}'=\bm{q} \bm{q}_R$), so long as one does not demand control over the global phase.  (The most general unitary 2x2 matrix has four real parameters, and the unit quaternion $\bm{q}_R$ has three; the missing parameter can be traced to the global phase of $\bm{q}'$.)

Since all 2x2 unitary operators $\bm{U}$ correspond to a rotation on the Bloch sphere, all quaternionic right multiplications also correspond to such a rotation.  Specifically, right-multiplying by $\bm{q}_R=exp(-\hat{\bm{n}} \gamma/2)$ is a rotation around the $\hat{n}$ axis by an angle $\gamma$, where $\hat{n}=f[\hat{\bm{n}}]$ as given by (\ref{eq:fdef}).  

Furthermore, couched in the language of quaternions, the state itself is effectively just another unitary operator.  In particular, the state $\bm{q}=exp(-\hat{\bm{n}} \gamma/2)$ is perhaps most naturally interpreted as a \textit{rotation} rather than a vector, a rotation that takes the z-axis ($\hat{z}$) to the state's standard vector representation on the Bloch sphere ($\hat{q}$).  There are many such rotations that will result in any given state vector, exactly corresponding to the many possible global phases of $\bm{q}$.

The global phase can be shifted by a quaternionic \textit{left} multiplication, $\bm{q}'=exp(\bm{i}\alpha) \bm{q}$.  Other left multiplications by unit quaternions $\bm{q}_L$ correspond to non-unitary operators.  In particular, any left-multiplication of the form shown in (\ref{eq:Tdef}) corresponds to the anti-unitary time-reversal operator $\bm{T}$.  Other non-unitary operators (which have found use in quantum information theory) are guaranteed not to change the normalization of $\bm{q}$ so long as $\bm{q}_L$ is also a unit quaternion.  Such operators take pure states to mixed states, because that the resulting state vector is as ill-defined/unknown as the global phase of $\bm{q}$.  

Besides simplifying the dynamic equations of a spin state in a magnetic field, the ability to implement anti-unitary transformations via left-multiplication offers a view of time-reversal compatible with a classical perspective.  Specifically, it becomes far more straightforward to reverse all of the time-odd parameters in the dynamical equations, making the time-symmetry more clearly evident.  Although there are many ways to implement time-reversal via a left-multiplication, all of them conform to the usual $\bm{T}^2=-1$ if the \textit{same} left-multiplication is applied twice.

\subsection{The possible importance of global phase}

The results summarized in Section 5.1 hold whether or not global phase is a mere gauge, but the status of global phase is important for the results discussed below.  Therefore, a short discussion of this topic seems appropriate here.

The global phase of a single-particle quantum state is either a choice of gauge or an unknown hidden variable.  Although most physicists have come down in favor of the former option, there is no experimental evidence either way.  Indeed, we cannot even probe down to the Compton scale at which these phases would fluctuate.  (For an electron, this phase frequency is $\omega_0=m_ec^2/\hbar$, and $\omega_0^{-1}\approx 10^{-21}\,sec$ is several orders of magntitude shorter than the shortest laser pulses.)   It is rare to even see this $exp(-i\omega_0 t)$ oscillation explicitly in quantum equations, because it is typically removed \textit{on the assumption} that global phases are irrelevant.

Of course, photons have lower-frequency oscillations than electrons, but this point only sheds further doubt on the notion that this phase is mere gauge.  In the classical limit, the global phase of an electromagnetic wave is indeed meaningful, and in the quantum limit phase issues are necessarily addressed via quantum field theory.  The failure of quantum-mechanical states to fully describe photons is arguably an indication that phases are a bit more important than quantum mechanics would have us believe.

Finally, note that even in the absence of a measurement on the time-scale of an oscillation, another way to probe oscillations is via a reference oscillator.  And of course, there is an enormous body of experimental evidence that such relative phases are indeed meaningful.  One simple explanation for this fact would be that single-particle states have a meaningful global phase, and such experiments are measuring relative values of this phase.  Unfortunately, this analysis is confounded by the tensor-product structure of multiparticle quantum states, making this point inconclusive.  Still, according the so-called $\psi$-epistemic approaches to quantum theory \cite{Spekkens}, this tensor product structure naturally arises for states of knowledge, not the underlying (hidden) states of reality.  And with the experimental fact of relative phase measurements requiring some underlying explanation, $\psi$-epistemic approaches (at least) might be more inclined to see phase as a hidden variable rather than mere gauge.

These arguments are certainly not conclusive, but if global phase \textit{could} be a hidden variable, it is certainly not advisable to immediately dismiss it up front.  And if one does not discard the phase, the quaternion form of spinors is arguably the best way to see how the phase is interrelated with the qubit.  (Indeed, the fact that these two can not be cleanly separated is another reason to keep the phase.)  The chief implication of this viewpoint is that it appears more natural to extend the dynamics, as outlined in Section 4.3; this issue will now be discussed in detail.

\subsection{Second-Order Qubits}

Sections 3 and 4 demonstrated that when the standard dynamical equations for a spin-1/2 state in a magnetic field are written in quaternionic form, the first-order equations reveal a broken symmetry.  Namely, one particular pure unit quaternion $\hat{\bm{v}}$ must be singled out from all others (for the map $M_i$ used in most of the above analysis, this corresponds to $\hat{\bm{v}}=\bm{i} $.)  

Another way to see this broken symmetry is via the Lagrangian that would generate the Schr\"odinger-Pauli equation.  For a spin-1/2 state $\ket{\chi}$, given an arbitrary Hamiltonian $\bm{H}$, the inner-product form of the corresponding Lagrangian \cite{LinckThesis} can be written as
\begin{equation}
\label{eq:ketL}
L_1(\ket{\chi},\ket{\dot{\chi}})=\bra{\chi}\bm{H}\ket{\chi}-\hbar\, Im\ip{\dot{\chi}}{\chi}.
\end{equation}

Taking the imaginary part of this last inner product may look reasonable in such a form, but the inner product structure looks quite unnatural when framed in terms of quaternions \cite{Adler}.  In quaternionic form, under the map $M_i$, the last term in (\ref{eq:ketL}) looks instead like $\hbar \, Re(\bm{i}\dot{\bar{\bm{q}}} \bm{q})$, where the special pure quaternion $\bm{i}$ makes an explicit appearance.

The ultimate source of this broken symmetry is the very $\bm{S}^3\to \bm{S}^2$ Hopf fibration procedure that motivated this paper.  There are an infinite number of ways to reduce a 3-sphere to a 2-sphere, each corresponding to a particular choice of $\hat{\bm{v}}$.  And crucially, this choice must be made before the Lagrangian can even be written down.  In other words, the symmetry that relates the possible different Hopf fibrations is not evident at the level of the Lagrangian or the dynamics of $\ket{\chi}$; it is only evident on the higher-level representation of $\bm{S}^3$, where the unit quaternions reside. 

Furthermore, this symmetry need not be broken at all.  So long as one is willing to extend the Lagrangian (and dynamics) to a second-order form, it becomes easy to write a harmonic-oscillator-like Lagrangian in terms of quaternions, without reference to any particular Hopf fibration:
\begin{equation}
\label{eq:2ndL}
L_2(\bm{q},\dot{\bm{q}})=\frac{1}{2} \left\{ |\dot{\bm{q}}+\bm{qb}|^2-\omega_0^2|\bm{q}|^2 \right\}.
\end{equation}
Remarkably, if $\bm{q}$ is constrained to remain a unit quaternion (which can be imposed via $L_2=0$), there is no obvious new physics implied by this higher-order form.  Of course, $\bm{q}(t)$ will have a richer structure, in that it will obey the second-order differential equation (\ref{eq:bkge}) and it will require more initial data to solve.  (Instead of merely an initial value of $\bm{q}$, it will require initial values of both $\bm{q}$ and $\dot{\bm{q}}$.)  But these additional parameters turn out to be equivalent to the original (arbitrary) choice of $\hat{\bm{v}}$, so \textit{whatever} they happen to be, the resulting dynamics can always be cast back into the first-order form (\ref{eq:vspe}).  The choice of map $M_v$ between spinors and quaternions is no longer an arbitrary choice, but is determined by the now-meaningful (and effectively hidden) parameter $\hat{\bm{v}}$.

As discussed in Section 4.3, this expanded dynamics does not encompass any new solutions that might be interpreted as antimatter, and therefore this is not a disguised form of the Dirac equation.  Looking at the level of the Lagrangians (\ref{eq:ketL}) and (\ref{eq:2ndL}), it seems this procedure is instead analogous to the extension from the (first-order) Dirac equation to the (second-order) Feynman-Gell-Mann equation, or ``second order fermions'' as they are known in quantum field theory \cite{SOF1,SOF2}.  The above proposal has no spatial component in the equation, merely spin; through this analogy, one might call the solutions to (\ref{eq:bkge}) ``second order qubits''.  

The most obvious implication of these second order qubits is that the hidden parameter space is much larger than a mere global phase.  Now it also includes the two free parameters in $\hat{\bm{v}}$.  Indeed, with 3 free hidden parameters now corresponding to the same point on the Bloch sphere, the hidden sector is now \textit{larger} than the measurable sector.  (If one considers the phase ``half-measureable'', via relative phase measurements, then this might fall under the umbrella of theories in which one can know exactly half of the ontological parameters, as in \cite{Spekkens}.)  A future publication will discuss potential uses of this large hidden variable space in the context of entangled qubits.  

A more immediate consequence of this formalism is that it strongly indicates that quantum states evolving like $exp(+i \omega_0 t)$ should not be interpreted as having less energy than standard $\exp(-i \omega_0 t)$ states, but instead exactly the same energy.  Classical physics is perfectly clear on this fact (there are no negative-energy-density classical fields), but the single-time-derivative form of (\ref{eq:sse}) has obscured the time-even nature of energy when it comes to quantum systems.  Second order (quaternionic) qubits, on the other hand, have a time evolution that goes like $\exp(-\hat{\bm{v}}\omega_0 t)$, demonstrating a smoothly continuous set of solutions that pass from $\hat{\bm{v}}=+\bm{i}$ to  $\hat{\bm{v}}=-\bm{i}$.  In order for the sign of the energy to flip, one of the intermediate solutions must either have zero energy or some strange discontinuity that would destroy the above symmetries.  

Finally, it is worth stressing that when going from first-order to second-order, despite the dramatic change in the Lagrangian and the dynamical equations, there are remarkably few physical consequences.  Given the $|\bm{q}|=1$ normalization constraint, there are no new spurious solutions that cannot be interpreted as a standard spin-1/2 state, no unusual dynamics that would lead to a new prediction.  In fact, even though this procedure was motivated by viewing the global phase as more than mere gauge, at this point there seems no reason why one could not view the entire hidden parameter sector (the phase plus $\hat{\bm{v}}$, or $\bm{u}$) as a new, larger gauge to be fixed.  Such a project is beyond the scope of this paper, but would be interesting to explore. 

\section{Conclusions}

Although it is standard practice to remove the global phase of a given spinor, there is no continuous way to do this to the space of \textit{all} spinors, even if one separately keeps track of a phase parameter $\alpha$.  This means that one cannot choose phase factors for all qubits that would vary continuously over the entire Bloch sphere.  \cite{Urbantke}

One possible reading of this topological fact is that the global phase of a spin-1/2 state should be treated as mere gauge, simply because it cannot be universally defined.  But following this logic, there should be no reason to use spinors at all; one would simply represent spin-1/2 states as points on a 2-sphere, and use SO(3) rather than SU(2).

The reason this is not done is because it would throw away valuable phase information; for example, the geometric phase accumulated by a precession around the Bloch sphere.  (Again, considering that these Berry phases are in fact measureable, it seems reckless to assume they are a meaningless gauge.)  

Instead, we argue that a cleaner approach is to not remove the global phase at \textit{any} stage of the analysis.  Given this, the most natural mathematical object to encode the state space of a spin-1/2 particle is a unit vector on a 3-sphere, or a unit quaternion.  Symmetries of the state space are then the same as the symmetries on the 3-sphere.

But from this starting point, it appears difficult to map a unit quaternion to a (well-defined) spin-state on the Bloch sphere without breaking these very symmetries.  Specifically, choosing one particular Hopf fibration is equivalent to choosing a special pure quaternion, which then makes the original phase look discontinuous over the reduced state space (perhaps encouraging one to again discard it).

Remarkably, there is an alternate path, that does not require breaking any symmetries or discarding any phases.  The essential idea is to expand the state space to include \textit{two} quaternions, orthogonal to each other on the 3-sphere.  (These two quaternions correspond to $\bm{q}$ and $\dot{\bm{q}}$ for second order qubits, and their orthogonality $Re(\bar{{\bm{q}}}\dot{\bm{q}})\!\!=\!\!0$ ensures the $|\bm{q}(t)|=1$ normalization is preserved.)  For a given $\bm{q}$, the allowed values of $\dot{\bm{q}}$ then lie on a $2$-sphere, and $\dot{\bm{q}}$ effectively encodes \textit{which} Hopf fibration one should use to map $\bm{q}$ to the Bloch sphere. 

After this map has been performed, the same dynamics on the Bloch sphere is recovered, no matter which particular $\dot{\bm{q}}$ generated the map in the first place.  From this perspective the possible values of $\dot{\bm{q}}$ might be seen as a enlarged gauge group.  But given the real second-order Lagrangian (\ref{eq:2ndL}) that naturally generates the equations of motion relating $\bm{q}$ and $\dot{\bm{q}}$, it would be a stretch to treat the former as ontological and the latter as a gauge.  An alternative viewpoint is that $\dot{\bm{q}}$ is effectively a hidden variable, one that may find uses in novel approaches to quantum foundations.

One minimal example of a new direction that may be inspired by such a viewpoint results from rewriting the traditional Bloch sphere vector $\hat{q}=f[\hat{\bm{q}}]$ in terms of the canonical momentum of the Lagrangian $\bm{p}=\partial L_2/\partial \bm{\dot{q}}$.  From (\ref{eq:2ndL}) one finds $\bm{p}=\overline{\bm{\dot{q}}+\bm{qb}}$; then using (\ref{eq:vqhat}) and the effective dynamical equation (\ref{eq:vspe}), it transpires that $\hat{\bm{q}}$ is simply $\bm{pq}/\omega_0$. In other words, in the second-order qubit framework, the traditional quantum state $\hat{q}=f[{\bm{pq}}/\omega_0]$ naturally encodes the very underlying phase-space product crucial to the ``old'' quantum theory, but does not encode which particular phase-space orbit is hidden in $\bm{q}(t)$ and $\bm{p}(t)$.    

Even without enlarging the state space to this extent, viewing spinors in quaternion form has other advantages, most notably a straightforward way to implement time-reversal via left-multiplication.  More general non-unitary transformations also become easily available, which may be of interest to the field of quantum information (as well as any foundational proposals in which pure states naturally become mixed, via some new non-unitary process).  Finally, note that this recasting of quantum states allows pure states to have the same mathematical structure as generic (but phaseless) unitary operators; these can both correspond to unit quaternions.  In quaternion form, then, a spin state is more naturally viewed as a \textit{rotation}; perhaps unsurprising, given that these states encode angular momentum, but an interesting perspective nonetheless.

\section*{Acknowledgements}
The authors would like to specifically thank Carlos Salazar-Lazaro for introducing KW to quaternions in this context, and also Rebecca Linck for tireless preliminary analysis.  Further thanks go to Nick Murphy and Jerry Finkelstein.
\\
\\
	
\section*{Appendix: Quaternions}

	A quaternion is a number with one real and three imaginary components $(\bm{i,j,k})$. First described by William Rowan Hamilton in 1843, quaternions obey the following rules:
\begin{subequations}
\begin{align}
\bm{i}^2 = \bm{j}^2 = \bm{k}^2 = \bm{ijk} = -1 \nonumber \\
\bm{ij} = \bm{k} \hspace{8mm} \bm{ji} = -\bm{k} \nonumber \\
\bm{jk} = \bm{i} \hspace{8mm} \bm{kj} = -\bm{i} \nonumber \\
\bm{ki} = \bm{j} \hspace{8mm} \bm{ik} = -\bm{j} \nonumber
\end{align}
\end{subequations}
Note that quaternions do not in general commute.

The conjugate of a quaternion, $\bar{\bm{q}}$, has the same real component as $\bm{q}$, but opposite signs for each of the three imaginary components.  When conjugating a product, one can use $\overline{\bm{pq}}=\bar{\bm{q}}\bar{\bm{p}}$, but it is crucial to change the multiplication order, just as in a Hermetian conjugate of a product of matrices.  A unit quaternion is defined as $|\bm{q}|^2 \equiv \bm{q}\bar{\bm{q}}=1$.  (The norm $|\bm{q}|$ is the square root of the real value $\bm{q}\bar{\bm{q}}$.)  Explicitly,
\begin{eqnarray}
\bm{q} = A + \bm{i}B + \bm{j}C + \bm{k}D  \nonumber \\
\bar{\bm{q}} = A - \bm{i}B - \bm{j}C - \bm{k}D  \nonumber \\ 
| \bm{q} |^2 =  \bm{q} \bar{\bm{q}}  = \bar{\bm{q}} \bm{q} =  A^2 + B^2 + C^2 + D^2. \nonumber
\end{eqnarray}

A pure quaternion has no real component.  Pure unit quaternions therefore have two free parameters, and can map to a unit vector on a 2-sphere.  In this paper, pure unit quaternions are generally notated as $\hat{\bm{v}}$.  (This is distinct from unit vectors on the Bloch sphere which are not bold; $\hat{v}$.)

Multiplying two unit quaternions always results in another unit quaternion, because $(\bm{pq})(\bm{\overline{pq}}) = \bm{pq\bar{q}\bar{p}}=\bm{p\bar{p}}=1$.  Note that terms of the form $\bm{u} \hat{\bm{v}} \bm{\bar{u}}$ do not have a real component, even if $Re(\bm{u})\ne 0$.  For example, if $\hat{\bm{v}}=\bm{i}$,
\begin{eqnarray}
\begin{aligned}
\bm{q} = \bm{ui\bar{u}} =& (A + \bm{i}B + \bm{j}C + \bm{k}D)\bm{i}( A - \bm{i}B - \bm{j}C - \bm{k}D) ,\nonumber \\
Re( \bm{q} ) =& -AB + AB -CD +CD = 0.
\end{aligned}
\end{eqnarray}
Thus, if $\bm{u}$ is a unit quaternion, $\bm{u} \hat{\bm{v}} \bm{\bar{u}}$ is guaranteed to be a \textit{pure} unit quaternion.

\subsection*{Exponential Quaternions}

	Euler's formula can be generalized to quaternions as
\begin{equation}
\label{eq:exp}
e^{\hat{\bm{v}}\theta} \equiv cos(\theta) +\hat{\bm{v}}sin(\theta) .
\end{equation}
  As long as $\hat{\bm{v}}$ is a pure unit quaternion, $exp({\hat{\bm{v}}\theta})$ will also be a \textit{unit} quaternion, but it will not be pure unless $cos(\theta)=0$.   When multiplying two exponentials together, in general one cannot simply add exponents.  This is best seen by expanding the exponentials using (\ref{eq:exp}).  The exception to this is when both are exponentials use the same $\bm{\hat{v}}$, in which case the angles are indeed additive.

Every unit quaternion can be written in the exponential form (\ref{eq:exp}).  Note that given the full 2-sphere of possible pure unit quaternions $\hat{\bm{v}}$, even if the angle $\theta$ is restricted as $-\pi<\theta\le\pi$, there are still two exponential forms that map to the same unit quaternion, as $exp[\hat{\bm{v}}\theta] = exp[-\hat{\bm{v}}(-\theta)]$. 

	The most general, norm-preserving transformation of a quaternion involves both a left and a right multiplication by unit quaternions, $\bm{q}' = e^{\bm{\hat{u}} \phi}\bm{q}e^{\bm{\hat{v}} \theta}$.  (One cannot get to any given $\bm{q}'$ via a left- or a right- multiplication alone.)  To invert this transformation, one does not interchange the left- and right- terms, but merely conjugates them; $\bm{q} = e^{\text{-}\bm{\hat{u}} \phi}\bm{q}'e^{\text{-}\bm{\hat{v}} \theta}$. 
	
	Pure unit quaternions $\hat{\bm{v}}$ can be easily mapped to a vector $\hat{v}$ on a unit 2-sphere, either using $\hat{v}=f[\hat{\bm{v}}]$ as defined in (\ref{eq:fdef}) or another invertible map.  Many of the results in this paper hinge on the fact that a rotation of $\hat{v}$ around some axis $\hat{n}$ by an angle $\theta$ can easily be effected by quaternionic multiplication.   The quaternion corresponding to the rotation axis is $\hat{\bm{n}}=f^{-1}[\hat{n}]$, and the rotation is equivalent to the multiplication $exp(\hat{\bm{n}}\theta/2)\, \hat{\bm{v}} \,exp(-\hat{\bm{n}}\theta/2)$.  Mapping the resulting pure quaternion back to the 2-sphere will reveal the rotated vector.  (If one imagines this rotation in the space of unit pure quaternions, no mapping is required.)

\subsection*{Pauli Matrices vs i,j,k}

The Pauli matrices are defined as follows:

\begin{eqnarray}
\begin{aligned}
 \sigma_x \hspace{1mm}= \begin{pmatrix}
                0 \hspace{3mm} 1 \\
                1 \hspace{3mm} 0 
        \end{pmatrix} \hspace{5mm}
 \sigma_y \hspace{1mm}&= \begin{pmatrix}
                \hspace{1mm} 0 \hspace{3mm} $-i$ \\
               $i$ \hspace{4mm} 0 
        \end{pmatrix} \hspace{5mm}
 \sigma_z \hspace{1mm}= \begin{pmatrix}
                1 \hspace{3mm} 0 \\
                0 \hspace{3mm} $-$1 
        \end{pmatrix}  \nonumber
\nonumber \\
\end{aligned}
\end{eqnarray}

\begin{equation}
\label{eq:vecs}
\vec{\sigma} = \sigma_x \hat{x} + \sigma_y \hat{y} + \sigma_z \hat{z} 
\end{equation}

An important property of the Pauli matrices is their relation to rotations, as demonstrated in Eqn (\ref{eq:Rn}).  But more relevant to this paper, is the quantity $-i\vec{\sigma}$.  These matrices, call them $u_n = -i\sigma_n$, obey the same algebra as the imaginary quaternions $\bm{i}$, $\bm{j}$, and $\bm{k}$.  From our convention defined via the map $M_i$, we have $u_x$ $\Leftrightarrow$ $\bm{k}$, $u_y$ $\Leftrightarrow$ -$\bm{j}$, and $u_z$ $\Leftrightarrow$ $\bm{i}$.  This gives rise to the equivalent commutation relations: 
\begin{eqnarray}
\begin{aligned}\
[u_x,u_y] = 2u_z \hspace{7mm} [u_y,u_z] &= 2u_x \hspace{7mm} [u_z,u_x] = 2u_y\nonumber \\
[k,\text{-}j] = 2i \hspace{10mm} [\text{-}j,i] &=2k \hspace{10mm} [i,k] = \text{-}2j \\
\end{aligned}
\end{eqnarray}
As noted in the first line of Table 1, the operation of $u_x$ on a spinor $\chi$ is equivalent to right-multiplication of $\bm{q}=M_i[\chi]$ by $\bm{-k}$.  (The map $M_i$ between $\chi$ and $\bm{q}$ is defined by (\ref{eq:qdef}).)  The same pattern holds in the table for $u_y$ $\Leftrightarrow$ $\bm{j}$, and $u_z$ $\Leftrightarrow$ $-\bm{i}$.

\end{document}